\documentclass[twocolumn,showpacs,prl,superscriptaddress,preprintnumbers]{revtex4}
\usepackage{amssymb}
\usepackage{amsmath}
\usepackage{graphicx}
\usepackage[normalem]{ulem}
\usepackage[dvips]{color}

\usepackage[normalem]{ulem} 
\usepackage[dvips]{color} 
\renewcommand\sout{\bgroup \color{red} \ULdepth=-.5ex \ULset}

\usepackage{CJK}

\begin{document}
\begin{CJK*}{GBK}{song}


\title{Symmetry-energy dependence of the dynamical dipole mode in
the Boltzmann-Uehling-Uhlenbeck model}

\author{S. Q. Ye}
\affiliation{Shanghai Institute of Applied Physics, Chinese
Academy of Sciences, Shanghai 201800, China} \affiliation{University of Chinese Academy of Sciences, Beijing 100049, China}
\author{X. Z. Cai}
\affiliation{Shanghai Institute of Applied Physics, Chinese
Academy of Sciences, Shanghai 201800, China}
\author{Y. G. Ma \footnote{Correspondence author. Email:  ygma@sinap.ac.cn }}
\affiliation{Shanghai Institute of Applied Physics, Chinese
Academy of Sciences, Shanghai 201800, China}
\author{W. Q. Shen}
\affiliation{Shanghai Institute of Applied Physics, Chinese
Academy of Sciences, Shanghai 201800, China}

\begin{abstract}
Using an isospin-dependent
Boltzmann-Uehling-Uhlenbeck (IBUU) transport model, we
have studied the connection between the symmetry energy and features of the dynamical dipole
mode in fusion reactions with charge-asymmetric entrance channel.
The yield and angular distribution of the prompt photon emission are
extracted by a bremsstrahlung approach. The experimental data
of $^{36}$Ar+$^{96}$Zr at 16 MeV/nucleon and $^{32}$S + $^{100}$Mo
at 9.3 MeV/nucleon are compared with  IBUU model
calculations, and the soft symmetry energy is found to describe the data
reasonably well.
\end{abstract}


\maketitle
\date{\today}

Heavy-ion collision using radioactive nuclei
(especially $N > Z$) provides a unique possibility for exploring
the structure of nuclei and
the interaction between nucleons. One of
related topic is the determination of the density dependence
of symmetry energy ($E_{sym}(\rho)$), which is 
important  in understanding the nuclear
equation of state (EOS) of
asymmetric nuclear matter as well as the properties of
astrophysical objects \cite{Steiner05,Baran05}.

Giant dipole resonance (GDR), a well-established collective
vibration of protons against neutrons in nuclei, can be thermally
excited in dissipative heavy-ion reactions \cite{Snover86}.
The existence of a
different GDR-like excitation mode, called pre-equilibrium GDR or
dynamical dipole mode, was also addressed experimentally
\cite{Flibotte96,Pierroutsakou03,Pierroutsakou05,Pierroutsakou09,Pierroutsakou_17,Martin08}
and theoretically
\cite{Chomaz93,Baran01,Baran01b,Simenel01,Wu10,Tao,Baran09,Papa05}.
During the early stage of charge-asymmetric heavy-ion collisions, a
large amplitude collective dipole oscillation can be triggered along
the symmetry axis of the strongly deformed composite system. This
oscillation can emit prompt dipole photons, called dynamical dipole
radiation, in addition to the photons originating from a thermal
excited GDR of the hot compound nucleus (CN). The gamma
production from such a pre-equilibrium evolution contains lots of
information about the early stage of collisions when the CN is still
in a highly deformed configuration. Especially, the excitation can
reflect the density dependence of the $E_{sym}(\rho)$ below
saturation in the fusion or deep-inelastic processes.

Several microscopic transport models, such as time-dependent
Hartree-Fock (TDHF) \cite{Simenel01}, Boltzmann-Nordheim-Vlasov
(BNV) \cite{Baran09}, constrained molecular dynamics (CoMD)
\cite{Papa05}, and isospin-dependent quantum molecular
dynamics (IQMD) \cite{Wu10,Tao}, have been successfully used
to study the properties of dynamical dipole emission. A
"bremsstrahlung" approach derived in Refs.~\cite{Baran01,Baran01b}
was widely adopted to estimate the contribution of this
pre-equilibrium collective dipole radiation emission.  Here we use
an isospin-dependent
Boltzmann-Uehling-Uhlenbeck (IBUU) transport model \cite{Ma12}
 to calculate the gamma yield and the
emission anisotropy of such prompt dipole radiation. The reactions
of $^{36}$Ar + $^{96}$Zr at 16 MeV/nucleon and $^{32}$S + $^{100}$Mo
at 9.3 MeV/nucleon,  for which the
experiments are well documented and the data
are available in Refs.
\cite{Pierroutsakou09,Martin08}, are considered
in the present study.

The BUU model \cite{Bauer86}  was widely used for
simulating heavy-ion collision dynamics and
extracting important information
of nuclear matter properties. The BUU equation reads 
\begin{align}
      & \frac{\partial f}{\partial t}+ v \cdot \nabla_r f - \nabla_r U
\cdot \nabla_p f  = \frac{4}{(2\pi)^3} \int d^3p_2 d^3p_3 d\Omega
\nonumber
\\ & \frac{d\sigma_{NN}}{d\Omega}v_{12}
 \times [f_3 f_4(1-f)(1-f_2) - f f_2(1-f_3)(1-f_4)] \nonumber
\\ & \delta^3(p+p_2-p_3-p_4),  \label{BUU}
                   \end{align}
where $\frac{d\sigma_{NN}}{d\Omega}$ and $v_{12}$ are the
in-medium nucleon-nucleon scattering cross section and
the relative velocity for the colliding nucleons,
respectively.  The isospin dependence was incorporated
into the model through the initialization and the nuclear mean
field potential $U$, which is parametrized as
\begin{equation}
  U(\rho,\tau_{z}) = a\left(\frac{\rho}{\rho_{0}}\right) +
  b\left(\frac{\rho}{\rho_{0}}\right)^{\sigma} + V_{asy}^{n(p)} (\rho,\delta), \label{Usk}
\end{equation}
where $\rho_0$ is the  nuclear 
saturation density and $\rho=\rho_n+\rho_p$ with
$\rho$, $\rho_n$, and $\rho_p$ being the nucleon,
neutron, and proton densities, respectively. The coefficients
$a$, $b$, and $\sigma$ are parameters for the isoscalar
EOS. Here we choose the parameters corresponding to the soft EOS
with the compressibility $K$ of 200 MeV ($a$ = -356 MeV, $b$ = 303
MeV, $\sigma$ = 7/6). The single particle symmetry
potential $V_{asy}$ has the form of 
\begin{eqnarray}
V_{asy}^{n(p)} (\rho,\delta) &=& \frac{\partial}{\partial
{\rho}}\left[\frac{C_{sym}}{2}
        \left(\frac{\rho}{\rho_{0}}\right)^{\gamma} \rho \delta^2\right] \nonumber \\
       &=& \frac{C_{sym}}{2}\left[(\gamma-1) \left(\frac{\rho}{\rho_{0}}\right)^{\gamma} \delta^2 {\pm}
       2 \left(\frac{\rho}{\rho_{0}}\right)^{\gamma} \delta \right], \label{Vsym}
\end{eqnarray}
where  $C_{sym}=35.19$ MeV is the symmetry
energy coefficient.
For simplicity, the $E_{sym}(\rho)$ with $\gamma$ = 0.5 and $\gamma$
= 2 in the present calculation is called as the soft and
the stiff symmetry energy, respectively, with the former
(latter) larger (smaller) at subsaturation densities but smaller
(larger) at suprasaturation densities. 

\begin{figure}
\begin{center}
\vspace{-.5cm}
\includegraphics*[scale=0.3]{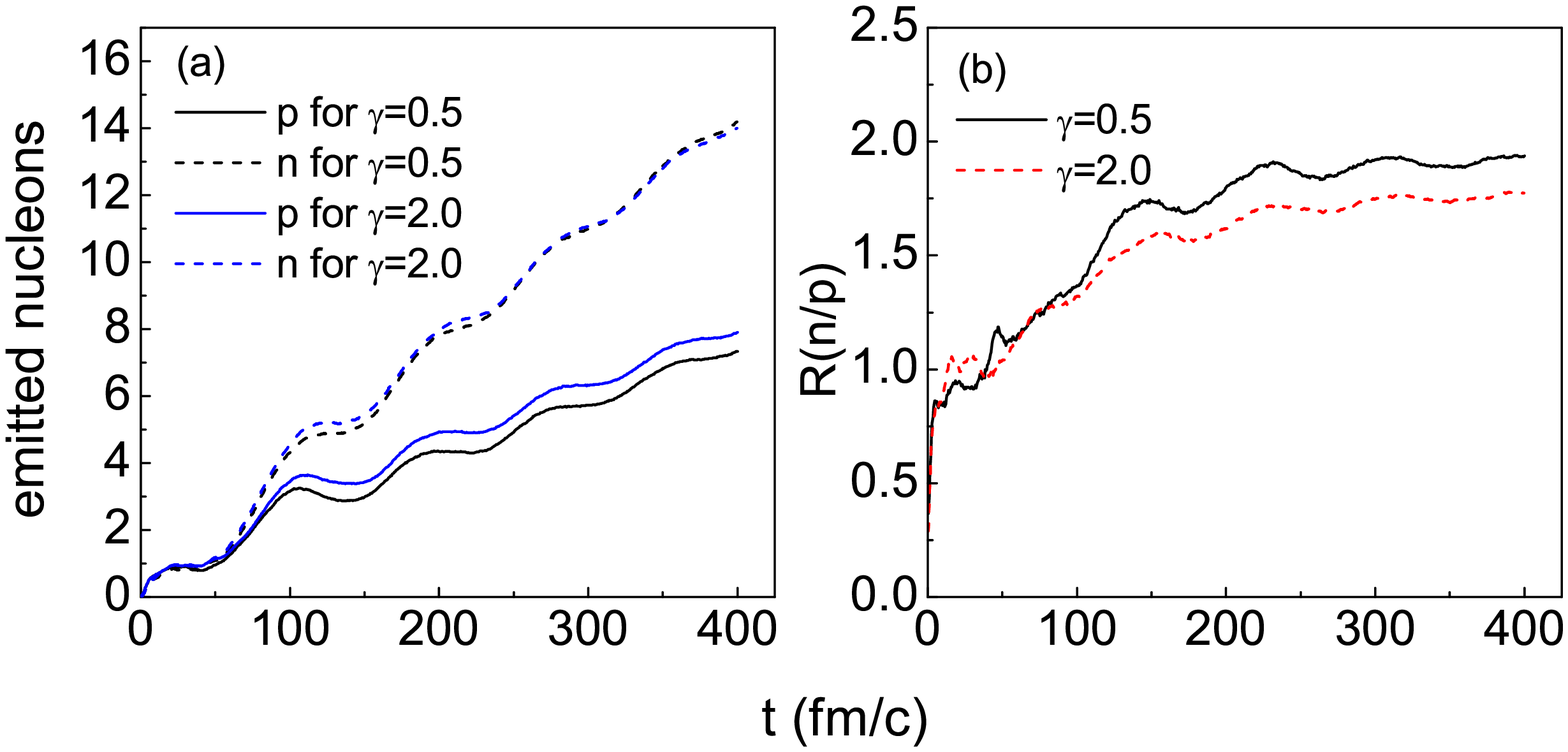}
\end{center}
\vspace{-2.8cm} \caption{(Color online) Time evolution of (a)
emitted neutrons and protons\protect; (b)
neutrons-to-protons ratio for $^{36}$Ar + $^{96}$Zr at 16
MeV/nucleon and \protect b = 0 fm.}
\label{fig-emission}
\end{figure}

As usual, the test particle method is applied in the IBUU
simulations, and the in-medium nucleon-nucleon cross section
is also taken into account \cite{Li93}. The calculations are
performed at impact parameters of  b = 0, 2,
and 4 fm where the fusion is an important reaction mechanism.
In the study of dynamic dipole
emission, the emitted nucleons  with their
local density below $\rho_0/8$ are excluded. Figure
\ref{fig-emission}(a) and (b) show the time evolution of the emitted
nucleons and the neutron-to-proton ratio ($R(n/p)$) 
from the soft and stiff symmetry energy, respectively, up to $t$ = 200 fm/c when about 10$\%$ of
the total nucleons become free. The stiff $E_{sym}(\rho)$ leads to
enhanced emission of protons resulting in a smaller $R(n/p)$. The
qualitative influence of $E_{sym}(\rho)$ on the $R(n/p)$ is similar
to  that predicted by the IQMD model \cite{Kumar}
in fragmentation.

During pre-equilibrium dipole oscillation, a larger initial dipole
moment will trigger a
larger-amplitude isovector oscillation, which increases the
chance of a clear experimental observation. The giant dipole moment
in coordinator space is defined as \cite{Baran01b}
\begin{equation}
D(t)=\frac{NZ}{A} X(t)=\frac{NZ}{A} (R_p-R_n),
\label{dt}
\end{equation}
where $A=N+Z$,  $N=N_p+N_t$, and $Z=Z_p+Z_t$
are the total number of participating nucleons, neutrons, and
protons from the projectile (p) and the target (t), respectively,
while $X(t)$ is the distance between the centers of mass of protons
and neutrons. The canonical
momentum conjugate of $D(t)$ is expressed as
\begin{equation}
DK(t)=\frac{NZ}{A} (\frac{P_p}{Z}-\frac{P_n}{N}), \label{dk}
\end{equation}
 where
$P_p$ and $P_n$ are the total momenta of
center-of-mass system for the protons
and neutrons. The initial 
($t$ = 0: touching configuration) giant dipole moment
can be expressed as \cite{Baran01}
\begin{equation}
D(t=0)=\frac{NZ}{A}X(t=0)=\frac{R_p+R_t}{N+Z}Z_{p}Z_{t}(\frac{N_p}{Z_p}-\frac{N_t}{Z_t}),
\label{dt0}
\end{equation}
where $R_p$ and $R_t$ are the radii of the projectile and target,
respectively.

\begin{figure}
\begin{center}
\includegraphics*[scale=0.33]{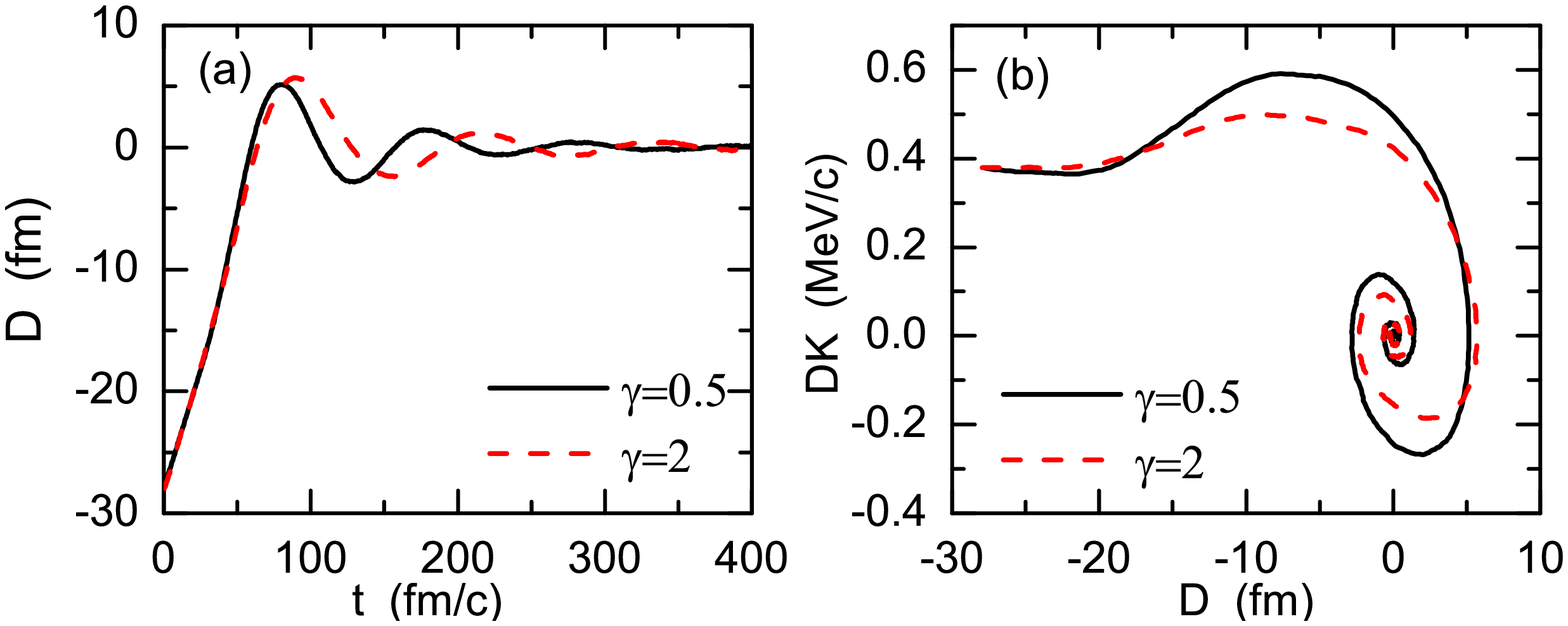}
\end{center}
\vspace{-.7cm} \caption{ (Color online) (a) Time evolution of dipole
moment \protect $D(t)$ in \protect
coordinate space; (b) dipole phase-space correlation
\protect for $^{36}$Ar + $^{96}$Zr at 16
MeV/nucleon and \protect b = 4 fm. The solid
(dashed) line represents for the result from the soft (stiff)
symmetry energy.} \label{dt-dk}
\end{figure}

In Fig. \ref{dt-dk} we plot the time evolution of the dipole moment
$D(t)$ and the phase  space correlation (spiraling) between $D(t)$
and $DK(t)$ for the $^{36}$Ar + $^{96}$Zr system at 16 MeV/nucleon
and $b$ = 4 fm.
It is clearly seen that the large amplitude of the first
oscillation decays rapidly within several periods.
The delayed dynamics for the stiff $E_{sym}(\rho)$ is related to the
weaker isovector restoring force. Especially, the right panel of
Fig. \ref{dt-dk} shows that the collective oscillation initiates at about 20 fm/c, corresponding to the touching configuration  of
the collision.

Finally, the gamma yield, as given by the bremsstrahlung approach,
can be extracted as \cite{Baran01,Baran01b}
\begin{equation}
\frac{dP}{dE_\gamma}=\frac{2e^2}{3 \pi \hbar c^3 E_\gamma}
\mid D''(\omega)\mid ^2,
\label{dpde}
\end{equation}
where $E_\gamma=\hbar \omega$ is the photon energy and $\mid
D''(\omega)\mid^2$ is the Fourier transform of the dipole
"acceleration" $ D''(\omega)=\int_{t_0}^{t_{max}} D''(t) e^{i \omega
t} dt$. For each event, $t_0$ represents the onset time
of the collective dipole response (phase-space spiraling), and
$t_{max}$ is the "damping time". In this work, after investigating
the $D(t)$ behavior, we define $t_{max}-t_0$ = 300 fm/c to make sure
that the collective oscillation is basically over.

\begin{figure}
\begin{center}
\includegraphics*[scale=0.33]{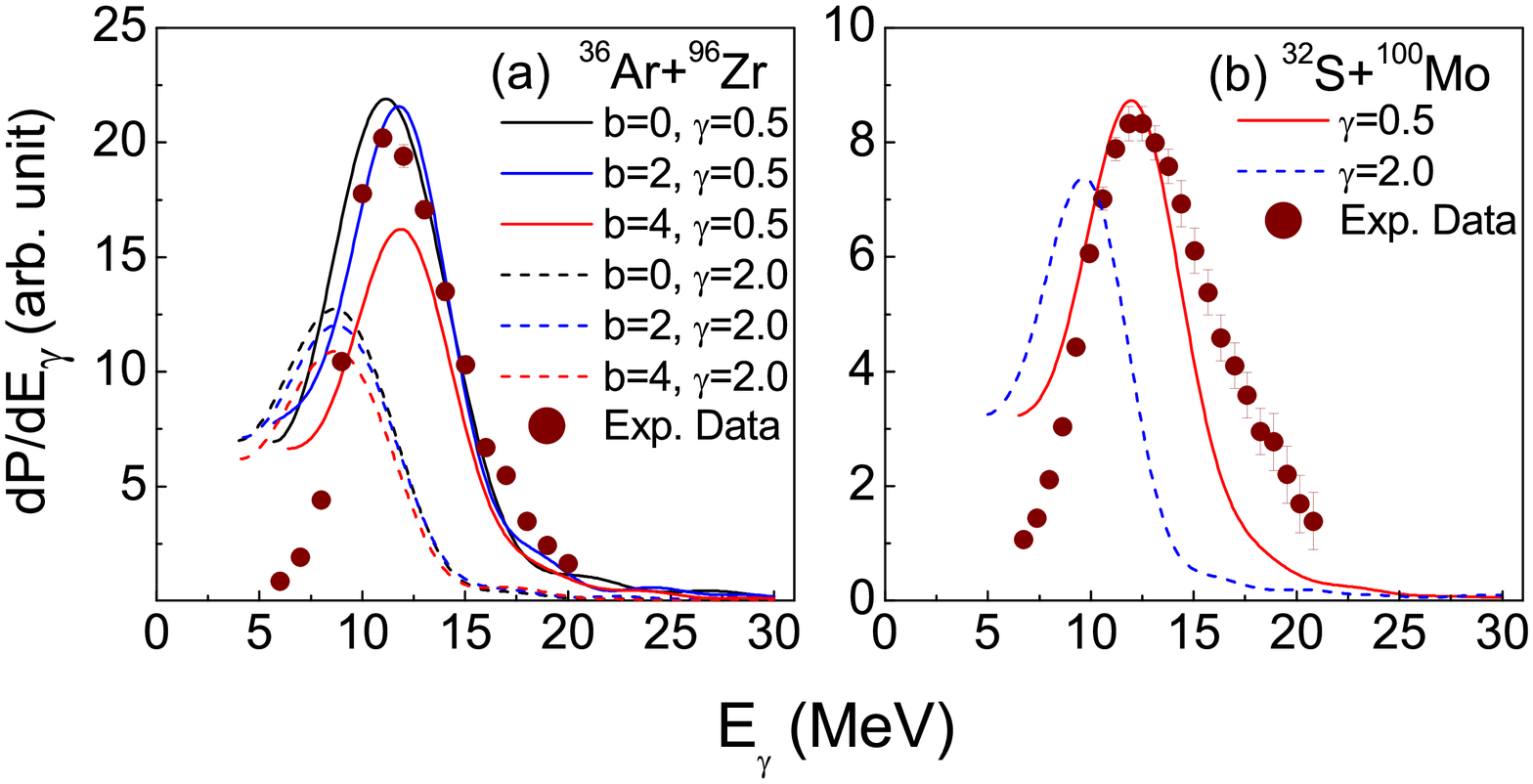}
\end{center}
\vspace{-2.8cm} \caption{ (Color online) Comparison of our
calculations with the data \protect of prompt photon
yields for $^{36}$Ar + $^{96}$Zr at 16 MeV/nucleon  (a) and $^{32}$S
+ $^{100}$Mo at 9.3 MeV/nucleon (b). } \label{dpde-E}
\end{figure}

Figure \ref{dpde-E} shows the comparison of calculated photon yields
with the experimental data of $^{36}$Ar+$^{96}$Zr at 16 MeV/nucleon
\cite{Pierroutsakou09} and $^{32}$S + $^{100}$Mo at 9.3 MeV/nucleon
\cite{Pierroutsakou_17}, respectively. It is seen that the soft $E_{sym}(\rho)$
 gives a
stronger restoring force for the dynamical dipole in dilute neck
region, and leads to a faster and  stronger
dynamical dipole oscillation,  resulting in a
larger centroid and stronger dynamical dipole emission. Another
feature of the dynamical dipole emission is that the gamma-ray
yields in both cases show a decreasing tendency towards larger
impact parameter inducing a  larger dipole moment
in the oscillation process.

\begin{figure}
\begin{center}
\includegraphics*[scale=.33]{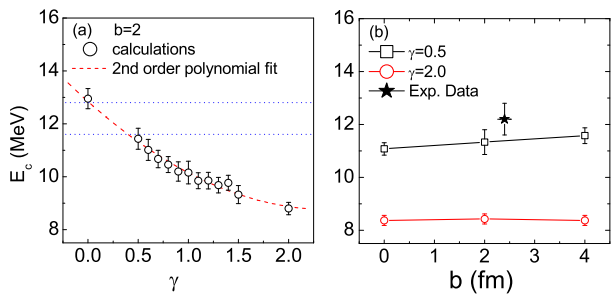}
\end{center}
\vspace{-3.cm} \caption{ (Color online) (a) \protect Dependence of the centroid
energy ($E_c$) on the symmetry energy factor $\gamma$ for
$^{36}$Ar + $^{96}$Zr at b = 2 fm. Two dashed lines represent the
upper and lower limit of the experimental data
\protect. (b) Comparison of $E_c$ \protect from the calculation of the soft
$E_{sym}$ ($\gamma$ = 0.5) and the stiff $E_{sym}$ ($\gamma$ = 2)
with the experimental data. } \label{dp-fit}
\end{figure}

To further check the systematic effect of symmetry energy
 factor ($\gamma$) on the centroid, we plot
the $\gamma$ dependence of the centroid energy $E_c$ in Fig.
\ref{dp-fit}(a). A monotonic decreasing behavior is exhibited in the
correlation between $E_c$ and $\gamma$. A constant symmetry energy
($\gamma$=0) gives the largest value of $E_c$. Because
the GDR takes place in the subsaturation density region, the
overall strength of the symmetry energy decreases with  increasing
$\gamma$. Figure \ref{dp-fit}(b) indicates that the soft symmetry
energy ($\gamma$ = 0.5) is in agreement with the experimental data
from Ref.~\cite{Pierroutsakou09}, where a mean
impact parameter  b$_{mean}$ = 2.3-2.5 fm was
deduced. However, we should  keep in mind that if
one considers momentum-dependent effective interaction which is
absent in the current IBUU model, the centroid and width of
dynamical dipole emission could be shifted \cite{Tao} and the above
conclusion might be modified.

Recently, a clear anisotropy for the angular distribution of the
total gamma spectrum  was observed
\cite{Martin08,Pierroutsakou09}. For the dipole oscillation just
along the beam axis, the angular distribution is expected to be
given by the Legendre polynomial expansion ${M(\theta)\sim \sin^2
\theta\sim 1+a_2P_2(\cos \theta)}$ with $a_2=-1$, where $\theta$ is
the polar angle between the emitted photon direction and the
oscillation axis. For the dynamical dipole mode, the prompt dipole
axis will rotate during the radiative emission. If the oscillation
is a stable and uniform rotation, an anisotropy parameter $a_2=-1/4$
is expected as averaging over all angles and orthogonalizing them
into the beam axis, see e.g., statistical
compound nucleus GDR radiation in Ref.\cite{Harakeh01}). The angular
distribution of the prompt gamma emission was extracted by
using the formalism from Ref. \cite{Baran09}.

\begin{figure}
\begin{center}
\includegraphics*[scale=0.21]{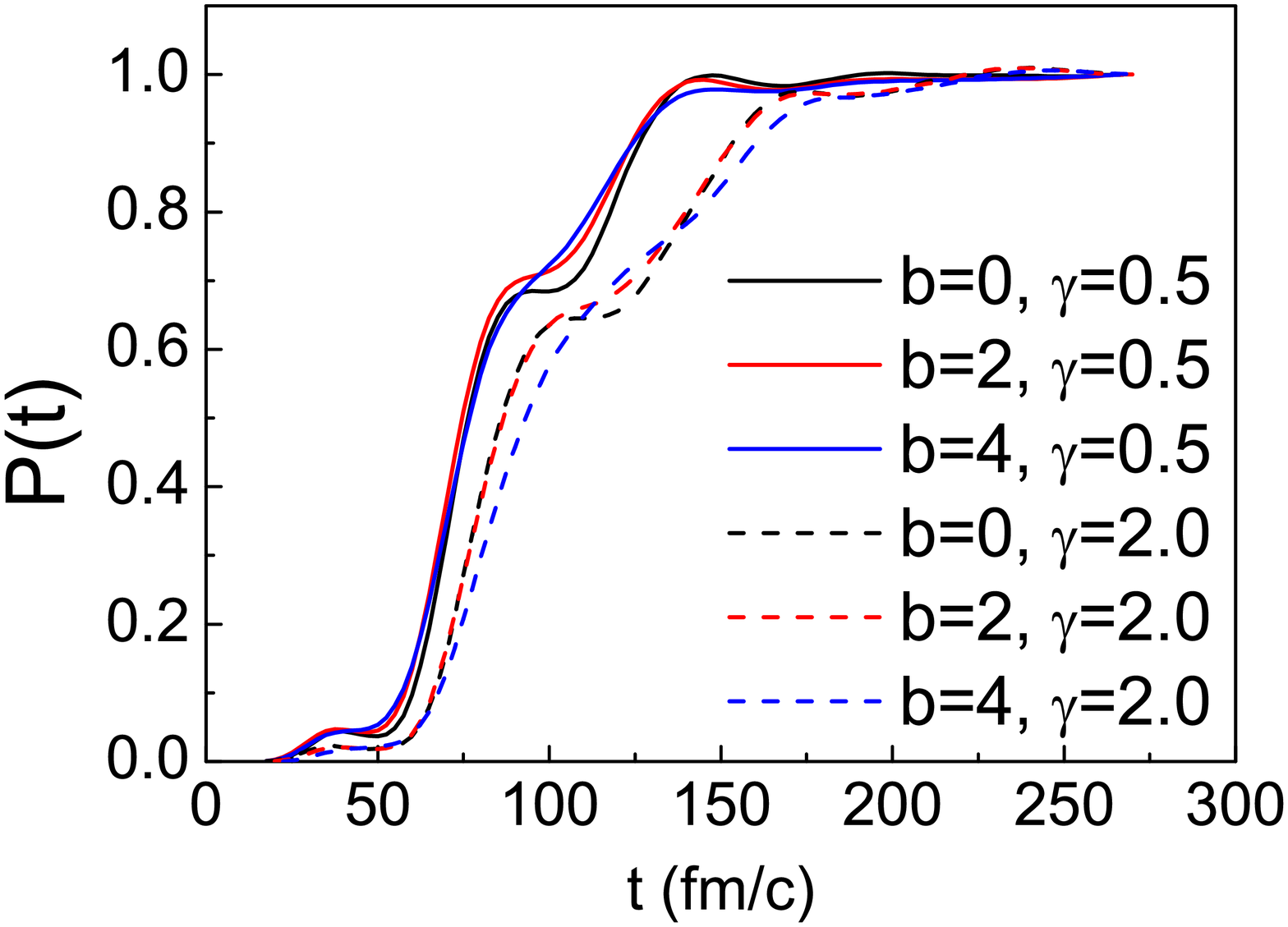}
\end{center}
\vspace{-.8cm} \caption{ (Color online) Time evolution of the
radiation emission probability for $^{36}$Ar+$^{96}$Zr at 16
MeV/nucleon with different impact parameters and $E_{sym}(\rho)$
\protect.  } \label{pt}
\end{figure}

First, we investigate in a very small time interval $\Delta t$. We
denote  $\phi_i$ and $\phi_f$ as the initial and final angles of the
oscillation axis with respect to the beam axis. So $\Delta
\phi=\phi_f-\phi_i$ is the rotation angle during the small time
interval $\Delta t$. From preceding text, the angular distribution
in the time interval $\Delta t$  can be averaged
over the angle $\Delta \phi$ as $
M(\theta)\sim1-(\frac{1}{4}+\frac{3}{4}x)P_2(\cos \theta),$
 where $x=\cos (\phi_f$+$\phi_i)\frac{\sin (\phi_f-\phi_i)}{\phi_f-\phi_i}$ \cite{Baran09}.
Then, for the whole time of the dynamical dipole, we can extract the
angular distribution by using a weighted form for every time step
and summing over all of them, i.e.,
 $M(\theta) = \sum_{i=1}^{t_{max}} \beta_i M(\theta,\Phi_i),$
where $\Phi_i$ is the rotation angle and $\beta_i$ is the radiation
emission probability.  From Eq.~(\ref{dpde}) we get
$\beta_i=P(t_i)-P(t_{i-1})$ where $P(t)=\int_{t_0}^{t} \mid D''(t)
\mid^2 dt / P_{tot}$ with $P_{tot}$ given by $P(t_{max})$, which is
the total emission probability at the final dynamical dipole damped
time. Finally we have:
\begin{equation}
 M(\theta) = M_0 \left[1-\sum_{i=1}^{t_{max}} \beta_i\left(\frac{1}{4}+\frac{3}{4}x\right)P_2(\cos \theta)\right],
 \label{m4}
\end{equation}
where $M_0$ can be obtained from the experimental data when
$P_2(\cos \theta)=0$. For the $^{36}$Ar + $^{96}$Zr system at 16
MeV/nucleon, $M_0 \approx 0.6\times10^{-4}$ 
from Ref. \cite{Pierroutsakou09}. To 
study the effect of the symmetry energy  on
the radiation emission probability, we plot the time evolution
of radiation emission probability  from different
symmetry energies  with different
impact parameters in Fig.~\ref{pt}. It is found that the photon
emission occurs before 200 fm/c, after which
 $P(t)$ levels off, and
the soft symmetry energy leads to a shorter
emission period. The line is smoother with a
larger impact parameter, because the rotation of the neck
region, i.e., the oscillation
axis, makes possible the radiation emission to
different angles in peripheral collision.

\begin{figure}
\begin{center}
\includegraphics*[scale=0.21]{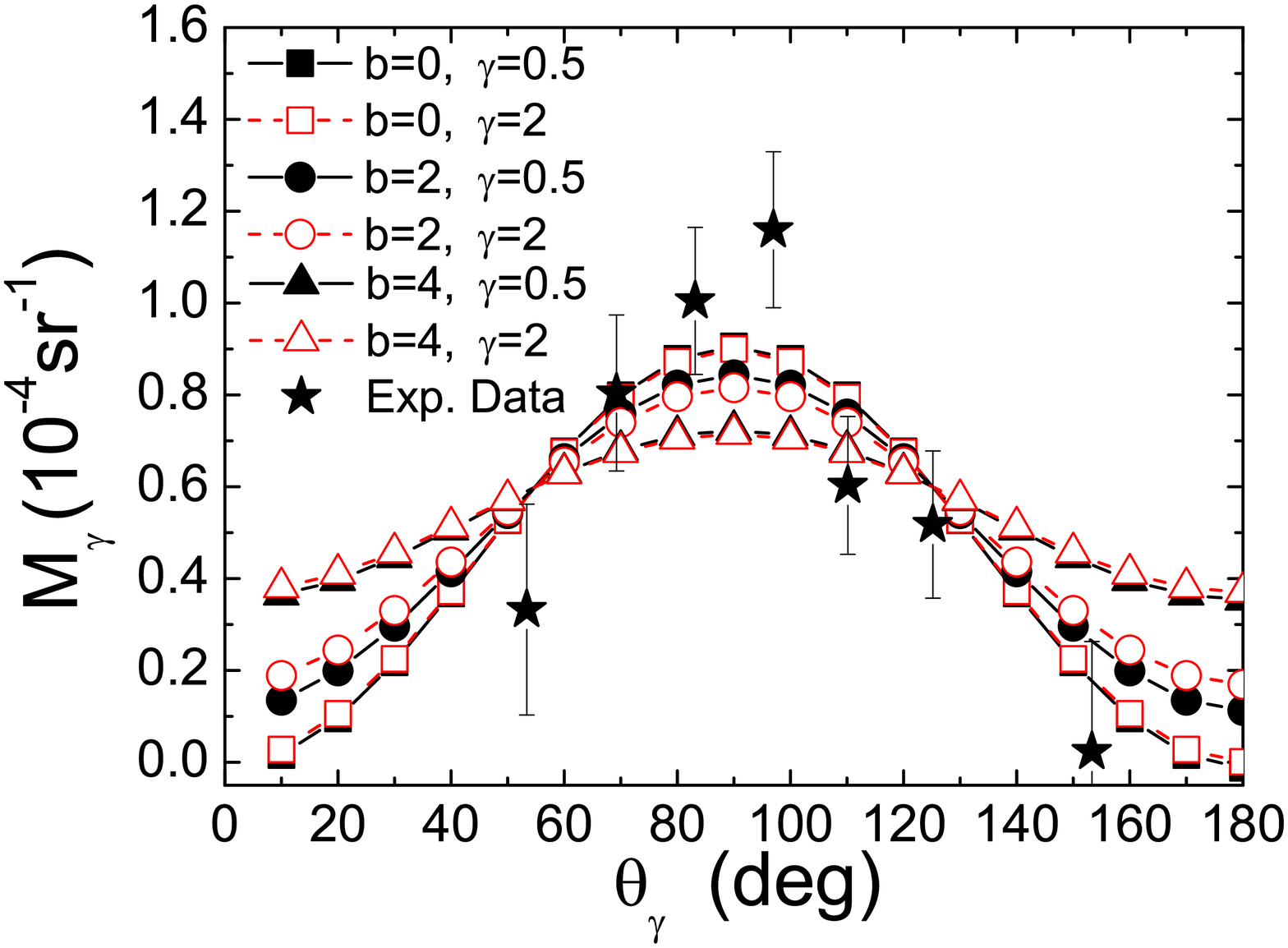}
\end{center}
\vspace{-.8cm} \caption{ (Color online)  Comparison of the
calculated angular distributions at different impact parameters with
the experimental data (solid stars) of $^{36}$Ar+$^{96}$Zr at
16 MeV/nucleon \protect. } \label{mtheta}
\end{figure}

The angular distribution of the dynamical dipole is shown in Fig.
\ref{mtheta} together with the experimental data from Ref.
\cite{Pierroutsakou09}. From different
symmetry energies with small
impact parameters, little rotation of the oscillation axis
leads to almost the same angular distribution with the coefficient
$a_2 = -1$, suggesting the
$\sin^{2}(\theta_\gamma)$ form of the emission from the dipole
oscillation along the beam axis. The experimental data from
Refs. \cite{Martin08,Pierroutsakou09} sustains our
calculation. In the non-central collisions a little more photons are
emitted at forward/backward angles for the stiff symmetry
energy, indicating that the angular
distribution of the dynamical dipole seems weakly dependent on
the symmetry energy  at subsaturation
densities.

In summary, the IBUU model has been applied to
study the properties of the dynamical dipole mode via a collective
bremsstrahlung mechanism in fusion reactions with charge-asymmetric
beams. We show the features of this collective mode for the  prompt
photon radiation with low-energy reactions of $^{36}$Ar + $^{96}$Zr
and $^{32}$S + $^{100}$Mo. The pre-equilibrium gamma-ray yield and
angular distribution are confronted with the experimental data. A
very similar analysis was also performed with BNV model in
Ref.~\cite{Pierroutsakou09}, where the integrated pre-equilibrium
gamma-ray yields were fitted with different symmetry
energies and nucleon-nucleon scattering
cross sections. The  centroid energy of the pre-equilibrium dipole
oscillation and its dependence on the symmetry energy are
studied in the present work. Our results demonstrate that centroid
energy of the dynamical dipole mode has a strong dependence on the
symmetry energy at subsaturation
densities. Similar results were also
found in Ref.~\cite{Baran09}, which provides a suitable probe to
test the symmetry energy as well as to investigate the early
entrance channel dynamics in dissipative reactions with asymmetric
nuclear collisions.  The present quantitative comparisons with the
dynamical dipole gamma-ray spectra data seem to favor a softer
 symmetry energy. However, a caution is
needed for the above conclusion of the symmetry energy if the
momentum dependent effective interaction is taken into account. On
the other hand, we hope new experiments could provide more accurate
angular distribution data for the prompt dipole radiation as
similarly expected in Ref. \cite{Baran09} for $^{132}$Sn beams,
which may give another constraint on the symmetry energy.

This work is supported partially by  the National Natural Science
Foundation of China   projects (No. 11220101005, 11035009,
10979074), the Major State Basic Research Development Program in
China (No. 2013CB834405 and 2014CB845401), and the Knowledge Innovation Project of
the Chinese Academy of Sciences under Grant No. KJCX2-EW-N01.

\end{CJK*}
\end{document}